\documentclass[aps,twocolumn,pra,superscriptaddress,amsmath,amssymb]{revtex4-2}
\usepackage{dcolumn}
\usepackage{bm}
\usepackage{amsmath}
\usepackage{txfonts}
\usepackage[T1]{fontenc}
\usepackage{xspace}
\usepackage{ulem}
\usepackage{comment}
\usepackage{braket}
\ifx\pdfoutput\undefined
\usepackage[dvipdfmx]{graphicx}
\usepackage[dvipdfmx]{hyperref}
\usepackage[dvipdfmx]{color}
\else
\usepackage{graphicx}
\usepackage{hyperref}
\usepackage{color}
\fi

\begin{document}
\let\emph\textit

\title{
  Hyperuniform properties of the square-triangle tilings
}

\author{Akihisa Koga}
\affiliation{
  Department of Physics, Tokyo Institute of Technology,
  Meguro, Tokyo 152-8551, Japan
}

\author{Shiro Sakai}
\affiliation{
  Center for Emergent Matter Science, RIKEN, Wako, Saitama 351-0198, Japan
}

\author{Yushu Matsushita}
\affiliation{
Toyota Physical and Chemical Research Institute, Nagakute, Aichi, 480-1192 Japan
}

\author{Tsutomu Ishimasa}
\affiliation{
  Division of Applied Physics, Graduate School of Engineering, Hokkaido University, Sapporo 060-8628, Japan
}

\date{\today}
\begin{abstract}
  We study hyperuniform properties for the square-triangle tilings.
  The tiling is generated by a local growth rule, 
  where squares or triangles are iteratively attached to its boundary.
  The introduction of the probability $p$ in the growth rule,
  which controls the expansion of square and triangle domains,  
  enables us to obtain various square-triangle random tilings systematically. 
  We analyze the degree of the regularity of the point configurations, 
  which are defined as the vertices on the square-triangle tilings,
  in terms of hyperuniformity.
  It is clarified that for $p<p_c \; (p_c\sim 0.5)$,
  the system can be regarded as a phase separation between square and triangular lattice domains and
  the variance of the point configurations
  obeys the scaling law $\sigma^2\sim O(R^{2-\alpha})$ with $\alpha<0$.
  The configurations are antihyperuniform.
  On the other hand, for $p>p_c$,
  the squares and triangles are spatially well mixed and
  the point configurations belong to the hyperuniform class III
  with the exponent $0<\alpha<1$.
  This means the existence of the hyperuniform-antihyperuniform transition
  at $p=p_c$.
  We also examine the structure factor of the square-triangle tilings.
  It is clarified that
  the peak structures in the large-wave-number regime are mostly common
  to all square-triangle tilings, while
  those in the small-wave-number regime strongly depend on 
  whether the point configurations are hyperuniform or antihyperuniform.
\end{abstract}

\maketitle

\section{Introduction}

Geometrical properties of the square-triangle tilings
have attracted much interest.
The square-triangle tiling is constructed
by densely packing the two-dimensional plane
with squares and triangles~\cite{Collins_1964,HenleyBook,Kawamura_1983,Oxborrow_Henley_1993,Xiao_2012,Clerc_2021}.
Both the periodic and quasiperiodic tilings have been observed
in various systems such as 
inorganic materials~\cite{exLadder1,exLadder2,Kageyama,Conrad_1998,Ishimasa_2011,Foerster_2013,Schenk_2017},
liquid crystals~\cite{Zeng_2004}, and
soft materials~\cite{Hayashida_2007}, as well as in numerical simulations~\cite{Louis_2012}.
In addition, 
tilings with nonperiodically distributed squares and triangles
have been observed in Mn-Cr-Ni-Si alloy~\cite{Ishimasa_2015}, 
mesoporous silica~\cite{Xiao_2012,Wang_2020}, 
and multicomponent block polymer~\cite{Matsushita}.
In this case, the configurations of these tiles are countless.
As a result, the square-triangle tilings are, in general, different from
the conventional periodic tiling, 
the deterministic quasiperiodic tilings, and the amorphous.
Therefore, a comprehensive characterization of the square-triangle tilings is desired.

Hyperuniformity is a framework to quantify a point distribution in a space~\cite{Torquato_2003,Torquato_2018}.
The point configuration is called hyperuniform if its variance in a large length scale
is smaller than a volume law.
The periodic and quasiperiodic point configurations are hyperuniform, and
their order metrics, which measure the degree of the regularity of the point configuration, 
have been examined~\cite{Torquato_2003,Torquato_2018,Zachary_2009,Lin_2017,KogaSakai}. 
Electronic properties on a quasiperiodic structure have also been discussed 
in terms of hyperuniformity~\cite{Fuchs,Sakai_2019,Sakai_2022}, 
which should be useful for characterizing the spatial distribution of 
the order parameter in broken-symmetry phases in the aperiodic systems~\cite{Jagannathan_1997,Wessel_2003,Koga_2017,Sakai_2017,Araujo_2019,Koga_2020,Sakai_Koga_2021,Ghadimi_2021,Inayoshi_2022,Koga_Coates_2022,SakaiB_2022,Matsubara_2024}.
Recently, it has been clarified that disordered point configurations in nature are also hyperuniform 
such as the distributions of colloid particles~\cite{Berthier_2011,Kurita_2011}
and avian photoreceptors~\cite{Jiao_2014},
amorphous silica~\cite{Zheng_2020}, 
vortex lattices in the superconductor~\cite{Llorens_2020}, 
marine algae system~\cite{Huang_2021}, and
galaxy cluster~\cite{Philcox_2023}.
Applications have been proposed such as
photonic and phononic crystals~\cite{Florescu_2009,Florescu_2009_2,Man_2013,Gkantzounis_2017}.
Then, a question arises:
are the square-triangle tilings hyperuniform or not?

The square-triangle tilings are hyperuniform
when the squares and triangles are arranged periodically and quasiperiodically.
The trellis and Shastry-Sutherland lattices,
which are also known as the $H$ and $\sigma$ phases in metallurgy,
are periodic and their point configurations are invariant under the two- and fourfold rotations.
These lattices are realized in $\rm SrCu_2O_3$~\cite{exLadder1,exLadder2}
and $\rm SrCu_2(BO_3)_2$~\cite{Kageyama}, 
where magnetic properties have been discussed~\cite{Gopalan,ShastrySutherland,MiyaharaUeda,Koga}.
There exist deterministic quasiperiodic tilings composed of squares and triangles~\cite{Stampfli,Hermisson_1997}.
The dodecagonal tiling has been observed
in realistic materials such as
Ta$_{1.6}$Te~\cite{Conrad_1998},
supramolecular dendritic liquid~\cite{Zeng_2004},
star-shaped terpolymers~\cite{Hayashida_2007}, 
and BaTiO$_3$ and SrTiO$_3$
on a Pt (111) substrate~\cite{Foerster_2013,Schenk_2017} .
In particular, the superconductivity has recently been observed in the compound Ta$_{1.6}$Te~\cite{Tokumoto},
which stimulates investigations on the two-dimensional dodecagonal quasicrystals~\cite{Ishimasa__2018,Schenk_2019,Yamada_2022}.
It has also been suggested that 
the tiling structure in Ta$_{1.6}$Te is away from
the ideal dodecagonal quasiperiodic configuration.
Furthermore, tilings with nonperiodically distributed squares and triangles have been observed,
as mentioned above.
In contrast to the periodic and quasiperiodic cases,
it is highly nontrivial whether or not such square-triangle tilings
are hyperuniform.

In this paper, we systematically study hyperuniform properties of the nonperiodic square-triangle tilings~\cite{Collins_1964,HenleyBook,Kawamura_1983,Oxborrow_Henley_1993,Xiao_2012,Clerc_2021}.
We here propose a local growth rule with a control parameter
to generate the tilings where the squares and triangles are
nonperiodically distributed.
Examining the variance and structure factor of the square-triangle tilings,
we demonstrate that random and hyperuniform tilings
are generated based on the local growth rule.
The relevance of realistic materials with dodecagonal symmetry is also addressed.

This paper is organized as follows.
In Sec.~\ref{rule},
we explain the square-triangle tiling and propose its growth rule.
Then, we clarify how tiling properties depend on the control parameter.
In Sec.~\ref{sec:hyper}, we study hyperuniform properties,
examining the variance and structure factor
of the square-triangle tilings.
The relevance of the materials with dodecagonal symmetry
is also addressed in Sec.~\ref{discussion}.
A summary is given in the last section.


\section{Square-triangle tiling}\label{rule}
\begin{figure}[htb]
  \includegraphics[width=0.8\linewidth]{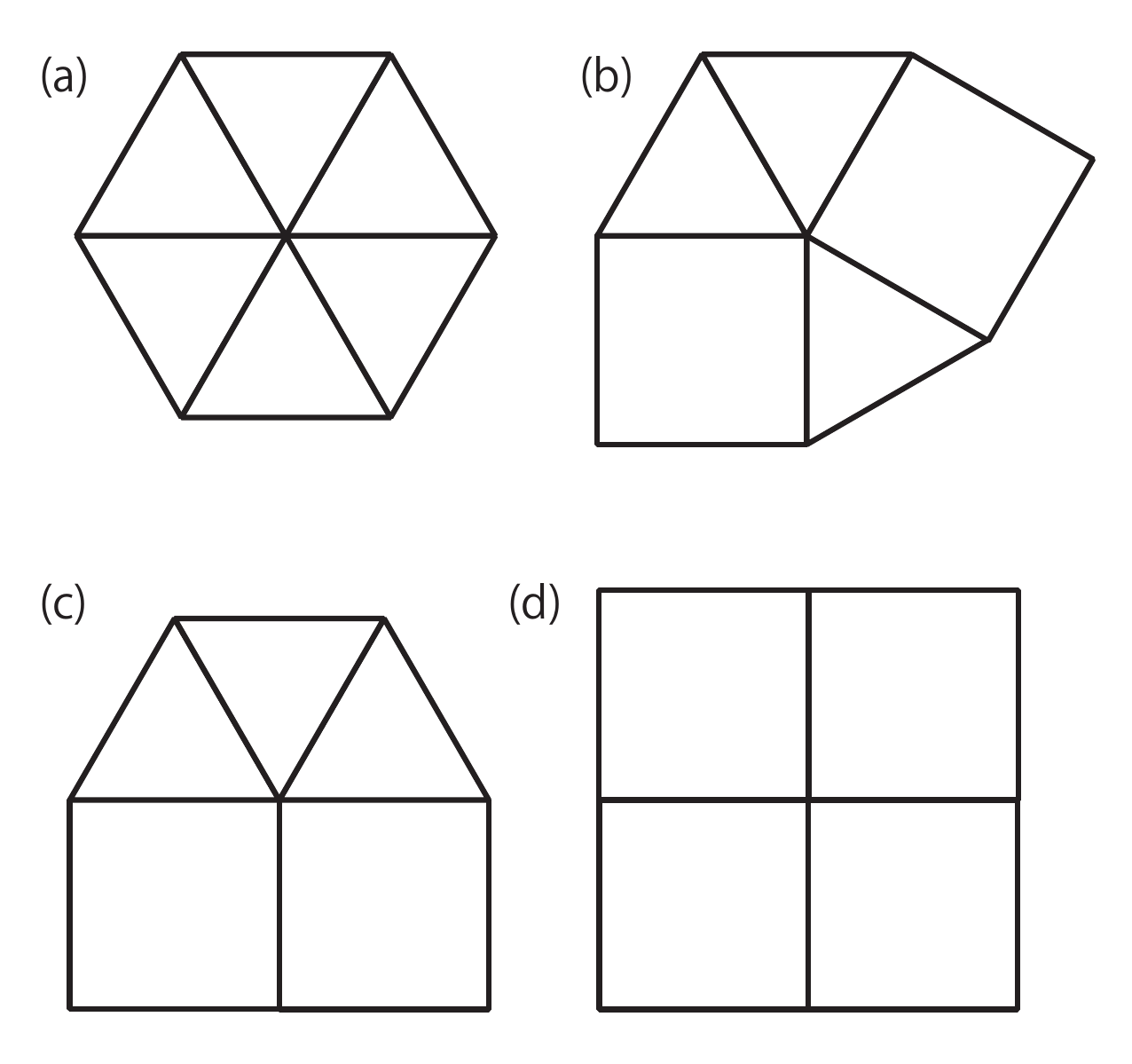}
  \caption{
    Four kinds of vertices in the square-triangle tiling.
    (a) $3^6$ vertex, (b) $3^2434$ vertex, (c) $3^34^2$ vertex,
    and (d) $4^4$ vertex.
  }
  \label{fig0}
\end{figure}

We consider the randomly distributed square-triangle tiling.
When the two-dimensional sheet is covered with squares and triangles,
the possible local configuration is restricted to four vertices, {\it i.e.}, 
$3^6$, $3^2434$, $3^34^2$, and $4^4$ vertices,
as shown in Fig.~\ref{fig0}.
\begin{figure}[htb]
  \includegraphics[width=0.8\linewidth]{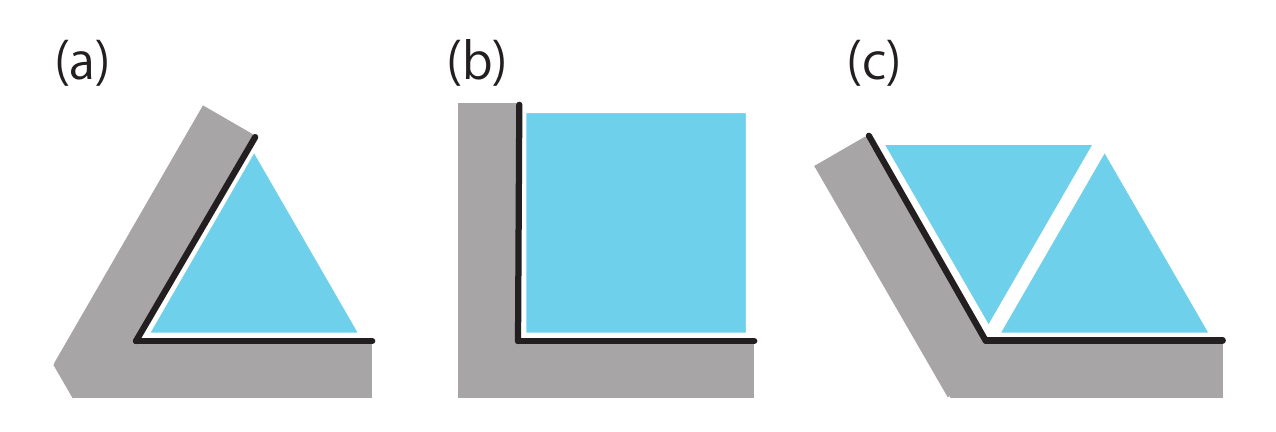}
  \caption{
    Growth rule for the vertices on gray shaded surface.
    The exterior angles are (a) $\theta=\pi/3$, (b) $\theta=\pi/2$,
    and (c) $\theta=2\pi/3$.
    Blue triangles and square can be uniquely attached.
  }
  \label{fig1}
\end{figure}
\begin{figure}[htb]
  \includegraphics[width=0.8\linewidth]{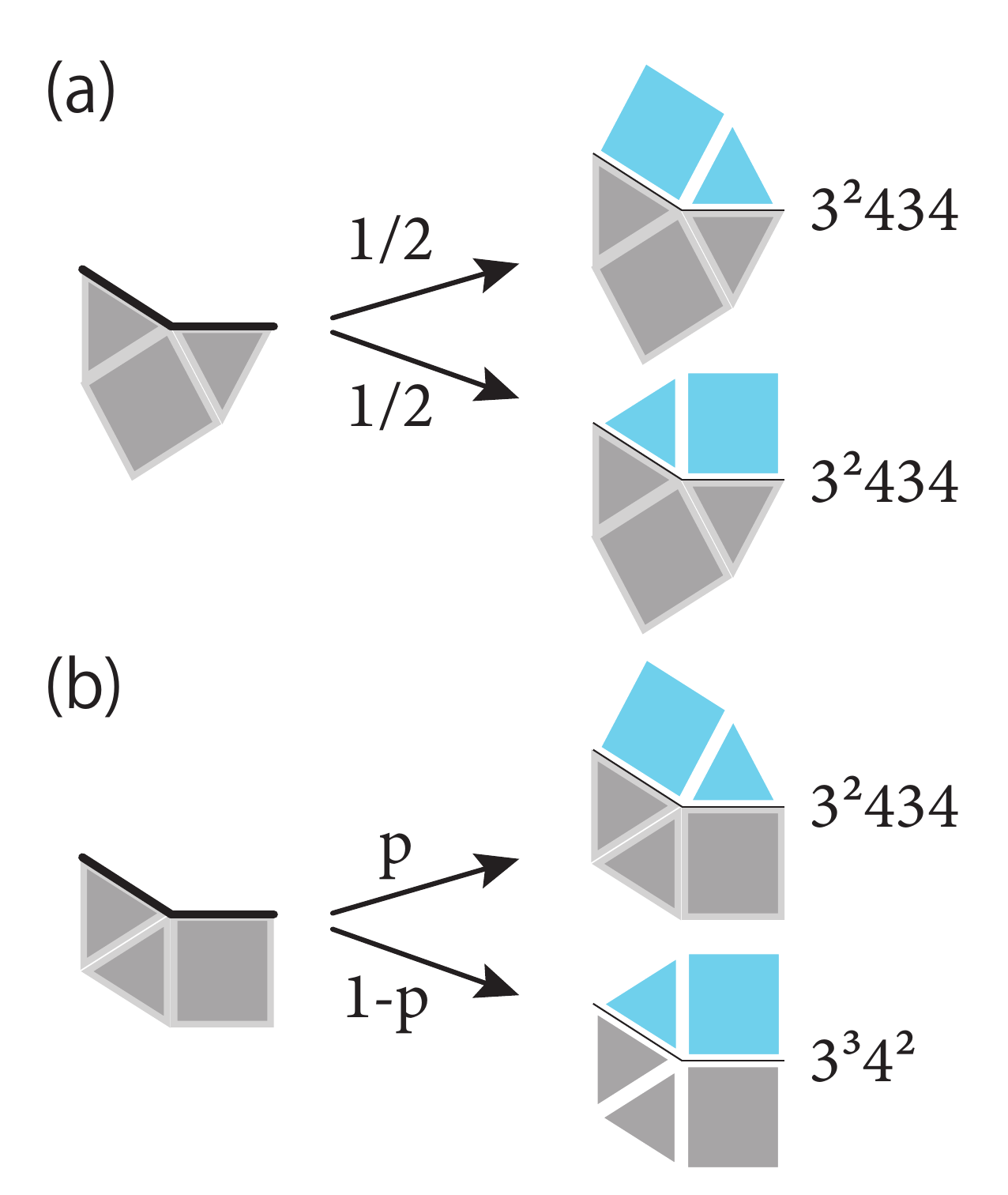}
  \caption{
    Growth rule on the vertex with $\theta=5\pi/6$.
    }
  \label{fig2}
\end{figure}
In this work, we propose the growth rule to generate the square-triangle tilings,
where squares and/or triangles are attached to the boundary of finite size tiling iteratively.
The growth rule we propose here is the following.
We consider the vertex on the boundary 
where the surface is formed by some adjacent squares and/or triangles.
Here, we focus on the exterior angle $\theta$ of the vertex.
When $\theta=\pi/3$
($\theta=\pi/2$) at a certain vertex,
a triangle (square) can be uniquely attached there,
which is shown as the blue triangle (square)
in Fig.~\ref{fig1}(a) [\ref{fig1}(b)].
Two triangles can be attached
when $\theta=2\pi/3$,
as shown in Fig.~\ref{fig1}(c).
When $\theta=5\pi/6$,
the adjacent square and triangle can be attached.
In contrast to the cases with $\theta=\pi/3, \pi/2$ and $2\pi/3$,
the arrangement of the square and triangle sharing a common edge
is not uniquely determined.
In this case, we choose it randomly and
attach the pair of tiles at the vertex.
When the surface structure is considered for the finite tiling,
there always exist the vertices with $\theta\le 5\pi/6$
except for the tiling being small.
Therefore, the above rules should be enough to grow the tiling.

We wish to note that when $\theta=5\pi/6$ at a certain vertex,
its surface structure is classified into two.
The corresponding internal structure is characterized by either $343$ or $3^2 4$,
which is schematically shown in Figs.~\ref{fig2}(a) and \ref{fig2}(b), respectively.
In the former case, the $3^2434$ vertex is always obtained
after the pair of tiles is attached.
Therefore, its arrangement is determined with equal probability.
By contrast, in the latter, either $3^2 434$ or $3^34^2$ vertex appears
after the square-triangle pair is attached.
This allows us to introduce the probability $p$.
Namely, the $3^2434$ vertex appears with $p$ and
the $3^34^2$ vertex appears with $1-p$.
An important point is that one can control
the expansion of the ``domain'' composed of either squares or triangles.
Namely, when $p$ is small (large),
the expansion of the domains should be enhanced (suppressed).
We note that when the tilings are generated by means of this growth rule, 
the number of random choices of the vertices with $\theta=5\pi/6$
is proportional to the total number of vertices.
This implies that the tilings are macroscopically degenerate,
in contrast to the Onoda-Steinhardt-DiVincenzo-Socolar rule
for the deterministic quasiperiodic tilings~\cite{OSDS}.
Some details to generate square-triangle tilings are shown in Appendix~\ref{G}.

Figure~\ref{fig:tilings} shows the square-triangle tilings
generated by the growth rule with $p=0.2$, 0.5 and 1.
\begin{figure*}[htb]
  \includegraphics[width=\textwidth]{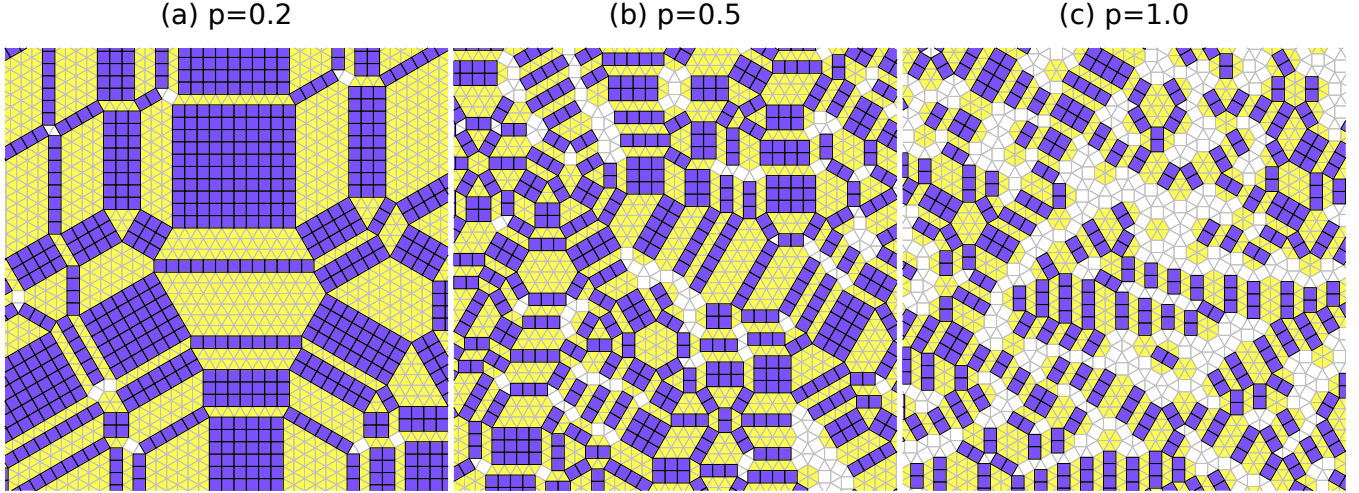}
  \caption{
    Square-triangle tilings generated by means of the growth rule with (a) $p=0.2$, (b) $p=0.5$, and (c) $p=1.0$. 
    Bold purple squares represent multiple square assembly more than two, 
    while yellow triangles express domains composed of more than three consecutive triangles.
  }
  \label{fig:tilings}
\end{figure*}
We find the large square and triangle domains
in the case with $p=0.2$.
With increasing $p$, such large domains disappear,
and squares and triangles tend to be mixed.
  To clarify local configurations in the tiling,
  we calculate the fractions of vertices and the ratio,
  examining 800 independent circular tilings with $R=300$.
  We show the fractions of the vertices in Fig.~\ref{fig:vertex}(a).
\begin{figure}[htb]
  \includegraphics[width=0.8\linewidth]{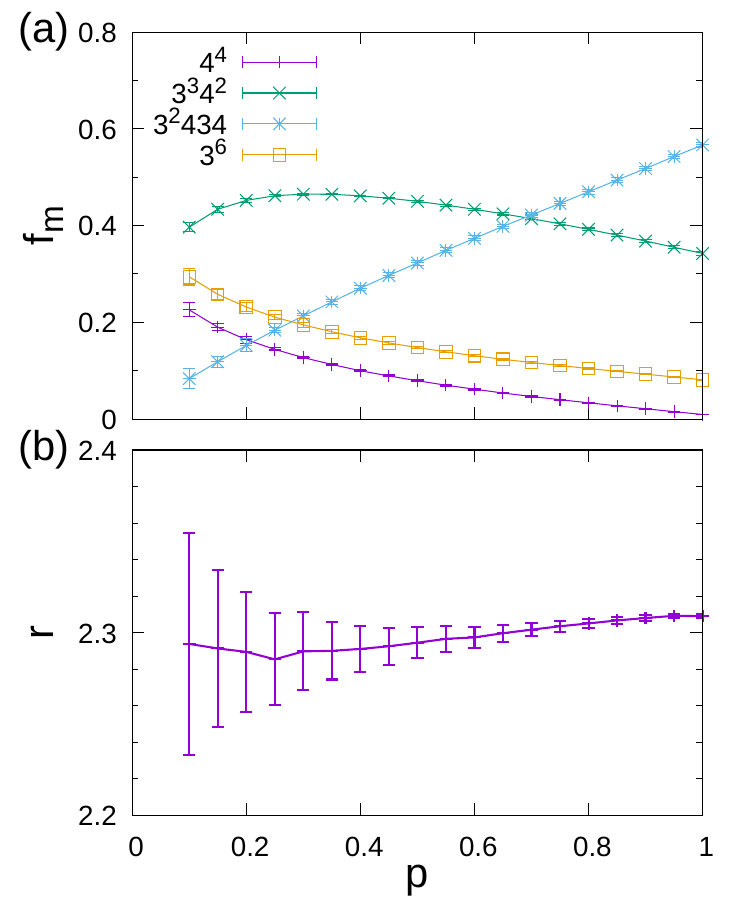}
  \caption{
    (a) Fractions of vertices and (b) ratio $r(=N_\triangle/N_\square)$
    as a function of $p$.
    The error bars represent the standard deviation of the results.
    }
  \label{fig:vertex}
\end{figure}
When $p$ is small enough, the two-dimensional sheet is almost covered with
the large square and triangle domains, and thereby
the fractions of $3^6$ and $4^4$ vertices are relatively larger.
As increasing $p$, the size of the domains becomes smaller, and 
$3^6$ and $4^4$ vertices appear with smaller probabilities.
Instead, the fractions of $3^34^2$ and $3^2434$ vertices increase.
The fraction of the $3^34^2$ vertex
has a maximum around $p\sim 0.3$ and decreases with increasing $p$.
By contrast, the monotonic increase appears in the $3^2434$ vertex.
This is due to the growth rule with $p$,
where the large $p$ suppresses to generate $3^34^2$ vertices.
Figure~\ref{fig:vertex}(b) shows the ratio of the number of triangles and squares.
We find that large fluctuations (error bars) appear when $p$ is small.
This originates from the fact that
large square and triangle domains appear in the tiling,
leading to the large sample dependence.
This is closely related to the variance of the point configurations,
which will be discussed later.
Nevertheless, we find that the ratio is almost 2.3,
which is similar to that of the deterministic Stampfli quasiperiodic tiling
with $4/\sqrt{3}\sim 2.309$.

\begin{figure}[htb]
  \includegraphics[width=\linewidth]{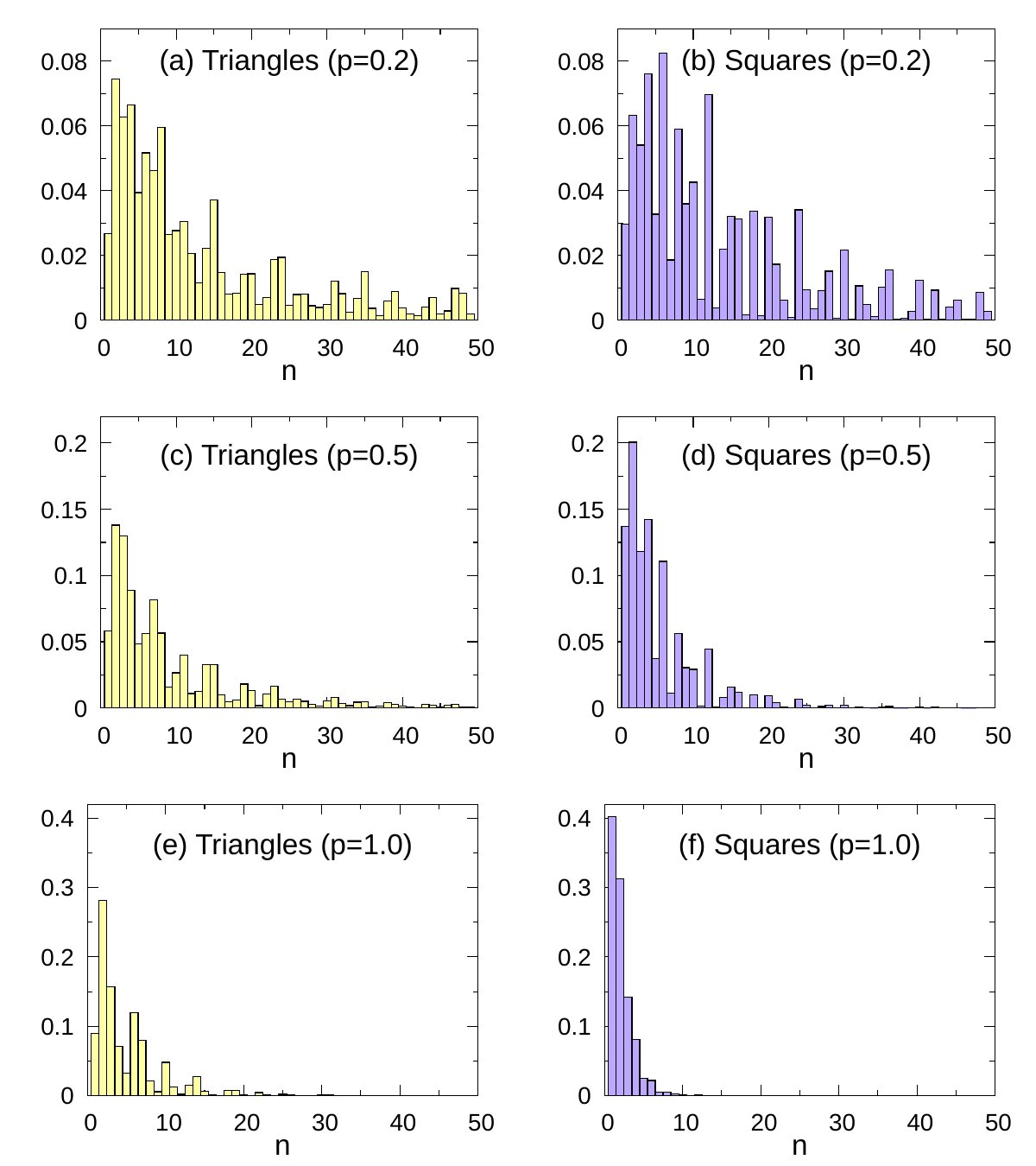}
  \caption{
  The tile fractions as a function of $n$, 
  where $n$ represents the number of triangles or squares occupying the domain.
  } 
  \label{vert}
\end{figure}
Figure~\ref{vert} shows the tile fractions in various size of square and triangle domains.
Few large domains appear in the case $p=1.0$, as shown in Figs.~\ref{vert}(e) and \ref{vert}(f).
On the other hand, larger fractions for triangle and square domains with large $n$ 
appear when $p=0.2$, 
as shown in Figs.~\ref{vert}(a) and \ref{vert}(b).
This means the coexistence of the large triangle and square domains in the tiling.

\begin{table}[htb]
  \caption{Vertex fractions $f_i \;(i=3^6, 3^2434, 3^34^2, 4^4)$,
    point density $\rho$, and
    ratio of numbers of triangles and squares $N_\triangle/N_\square$
    for periodic and quasiperiodic tilings composed of squares and triangles.}
  {
  \renewcommand\arraystretch{1.2}
  \begin{tabular}{c|cccc|cc}
    \toprule
    Lattice & $f_{3^6}$ & $f_{3^2434}$ & $f_{3^34^2}$ & $f_{4^4}$ & $\rho$ & $N_\triangle/N_\square$\\
    \hline
    Triangular & 1 & 0 & 0 & 0 & $1.155$ & $\infty$ \\
    Shastry-Sutherland & 0 & 1 & 0 & 0 & $1.072$ & 2\\
    Trellis & 0 & 0 & 1 & 0 & $1.072$ & 2 \\
    Square & 0 & 0 & 0 & 1 & 1 & 0\\
    \hline
    Stampfli & 0.071 & 0.746 & 0.182 & 0 & $1.077$ & 2.309\\
    \hline
    Our tiling ($p=1$) & 0.08 & 0.57 & 0.34 & 0.01 & 1.077 &2.3\\
    \toprule
  \end{tabular}
  }
  \label{tbl}
\end{table}
For comparison, we show in Table~\ref{tbl}
the fractions of vertices, point density $\rho$, the ratio
of the number of squares and triangles
for periodic and quasiperiodic point configurations.
Triangular, Shastry-Sutherland, trellis, and square lattices are composed of
squares and/or triangles, and
are composed of
only $3^6$, $3^2434$, $3^44^2$, and $4^4$ vertices, respectively.
In the deterministic Stampfli quasiperiodic tiling~\cite{Stampfli,Hermisson_1997},
$3^6$, $3^2434$, and $3^34^2$ vertices have finite fractions
while no $4^4$ vertices appear.
It is known that the above periodic and quasiperiodic tilings
belong to the hyperuniform class I and
their order metrics have been
obtained~\cite{Torquato_2003,Zachary_2009,Lin_2017,KogaSakai}.
In the randomly distributed square-triangle tiling with $p=1$,
the point density and ratio of the number of triangles and squares
are similar to those of the Stampfli quasiperiodic tiling.
By contrast, the fractions of four vertices are different.
This difference originates from the fact that, even when $p=1$,
the square-triangle tiling is generated
by randomly choosing the arrangement of the adjacent square and triangle
at the 343 vertex on the surface [see Fig.~\ref{fig2}(b)],
in contrast to the deterministic one.
The vertex structures in the perpendicular space is discussed 
in Appendix~\ref{Ap:perp}.
In the following section, we discuss hyperuniform properties
in the square-triangle tiling.

\section{Hyperuniform properties}\label{sec:hyper}
In this section,
we consider the point configurations, which are defined as the vertices 
on the square-triangle tilings.
Then, we discuss how these point configurations are characterized
in the framework of the hyperuniformity~\cite{Torquato_2003,Torquato_2018},
examining their variance and structure factor.

First, we focus on the variance of the point configurations,
which is defined as,
\begin{eqnarray}
  \sigma^2(R)=\langle N_{\bf r}(R)^2\rangle-\langle N_{\bf r}(R)\rangle^2,
\end{eqnarray}
where $N_{\bf r}(R)$ is the number of points in the circular region
with center coordinate ${\bf r}$ and radius $R$, and
$\langle N_{\bf r} \rangle$ is the average of $N_{\bf r}$
for many coordinates ${\bf r}$.
It is known that the point configurations are classified
according to the global asymptotic growth rate of the variance.
When the variance is proportional to the volume for large $R$ ($\sigma^2\sim R^2$),
the point configuration is regarded to be randomly distributed.
The point configuration is hyperuniform when
the variance obeys a scaling law smaller than the volume law.
According to the paper~\cite{Torquato_2018},
the hyperuniform point configurations in two dimensions are furthermore classified into
three groups:
\begin{equation}
  \sigma^2(R)\sim\left\{
  \begin{array}{ll}
    R & {\rm class\; I}\\
    R\log R & {\rm class\; II}\\
    R^{2-\alpha} & {\rm class\; III}
  \end{array}
  \right.,
\end{equation}
with $0<\alpha<1$.

\begin{figure}[htb]
  \includegraphics[width=\linewidth]{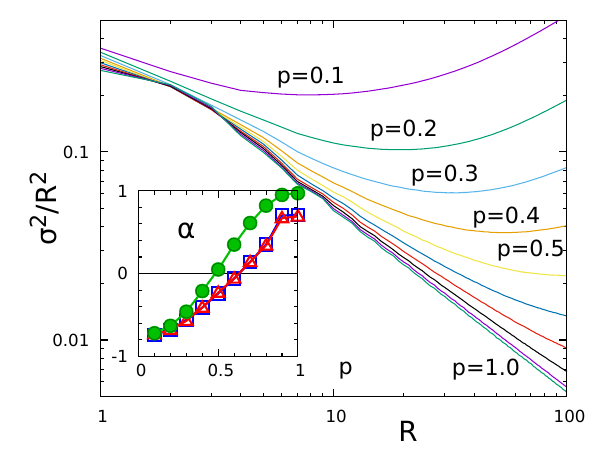}
  \caption{
    Double logarithmic plot of the quantity $\sigma^2/R^2$
    as a function of $R$.
    Inset shows the exponent of the variance in the large $R$ region.
    Solid circles represent the results for the point configurations for the square-triangle tilings.
    Open squares and triangles represent those for the S and T point configurations (see text).
  }
  \label{fig:var}
\end{figure}

Now, we evaluate the variance $\sigma^2(R)$ for the square-triangle tilings
with several $p$.
Figure~\ref{fig:var} shows the double logarithmic plot of $\sigma^2/R^2$, 
  by examining more than two thousands circular tilings.
It is found that the variance in the large $R$ region
strongly depends on the probability $p$.
For $p<p_c(\sim 0.5)$,
with increasing $R$, $\sigma^2/R^2$ once decreases, takes a minimum
at a certain $R$, and at last increases.
In the large $R$ region,
the variance is proportional to $R^{2-\alpha}$ with $\alpha<0$.
This should originate from the following.
In the small $p$ case, there exist large square and triangle domains
in the tiling, as shown in Fig.~\ref{fig:tilings}(a).
Therefore, the tiling can be regarded as the phase separation
of square and triangular lattices, and
the variance of the point configuration grows beyond the volume law.
On the other hand, when $p>p_c$,
$\sigma^2/R^2$ monotonically decreases, as shown in Fig.~\ref{fig:var}.
To clarify the asymptotic rate, 
we deduce the exponent $\alpha$ of the variance in the large $R$ region, 
which is shown in the inset of Fig.~\ref{fig:var}.
The increase of $p$ monotonically increases $\alpha$.
For $p>p_c$, $0<\alpha<1$, meaning that
the point configuration belongs to the hyperuniform class III. 
When $p\gtrsim 0.9$, the exponent approaches unity, 
meaning that the point configuration approaches the hyperuniform class I.
This suggests that the randomly distributed square-triangle tiling
with $p\sim 1$ has similar properties
to the deterministic Stampfli quasiperiodic tiling of the hyperuniform class I.

If one regards the center, instead of the vertices, of each square (triangle) as the point,
an S (T) point configuration is obtained.
The exponents of the variances of the S and T point configurations
are shown as open squares and open triangles in the inset of Fig.~\ref{fig:var}.
The variances are different from each other (not shown),
but the two exponents are almost the same.
Furthermore, these are smaller than the exponents of the vertex-point configuration.
Therefore, the hyperuniform-antihyperuniform transition 
in the S and T point configurations occurs at 
a value $(p\sim 0.6)$ which is distinct from that of the vertex-point configuration.

To clarify what happens at the transition point $p=p_c$, 
we now examine the structure factor, which is given as
\begin{eqnarray}
  S({\bf q})&=&\frac{1}{N}\sum_{ij}\exp\left[i{\bf q}\cdot({\bf r}_i-{\bf r}_j)\right],
\end{eqnarray}
where ${\bf r}_i$ is the coordinate for the $i$th vertex.
This quantity corresponds to the intensity
in the case of diffraction experiments to characterize the solid.
As for hyperuniform properties, 
the structure factor for $q\sim 0$ is classified as,
\begin{equation}
  S(q)\sim
  \begin{array}{l}
    q^\alpha \;\;\; \left\{
    \begin{array}{ll}
      \alpha> 1 & {\rm class\; I} \\
      \alpha=1 & {\rm class\; II} \\
      0<\alpha<1 & {\rm class\; III} 
    \end{array}
    \right.,
  \end{array}
\end{equation}
and $S(q\rightarrow 0)$ is nonzero for nonhyperuniform point configuration.
Note that $S(q\rightarrow 0)\neq S(q=0)=N$.
The divergence at $q=0$ originates from the forward scattering
in the thermodynamic limit,
which plays no role in hyperuniform properties.
To exclude this, it is convenient to introduce
the two-point correlation function $g({\bf r})$.
This is represented by
the histogram $\{\omega_k, r_k\}$ as
\begin{eqnarray}
g({\bf r})&=&\frac{1}{\rho}\sum_{ij}\delta\left({\bf r}-({\bf r}_i-{\bf r}_j)\right)\\
g(r)&=&\frac{1}{\rho}\sum_{r_k<R_{max}}\frac{\omega_k}{2\pi r_k}\delta(r-r_k),
\end{eqnarray}
where $\omega_k$ is the weight of the two-point distance $r_k$ in the tiling and 
$R_{max}$ is the upper limit of the histogram.
Then, the structure factor $S({\bf q})$ is represented as,
\begin{eqnarray}
  S({\bf q})&=& 1+\rho \int e^{-i{\bf q}\cdot{\bf r}}\left[ g({\bf r})-1 \right]d{\bf r} ,\\
  S(q)&=&1+2\pi\rho\int r J_0(qr)\left[ g(r)-1 \right]dr\\  
    &=&1+\sum_k \omega_k J_0(qr_k) -\frac{2\pi R\rho}{q}J_1(qR_{max}),\label{eq:sq}
\end{eqnarray}
where $J_n$ is the Bessel function of the first kind.
Two point correlations can be correctly taken into account
in the histogram, and
$R_{max}$ leads to oscillation behavior
in the structure factor.
To suppress the oscillation, it is useful to use the Gaussian filter.
The details are shown in Appendix~\ref{A}.
In this calculation, we obtain the structure factor,
combining the histograms with $R_{max}=100$ and
the Gaussian filter with $\sigma=0.07$.
In this case, unavoidable effects due to the width of Gaussian appear
in the tiny region $q<q_c$ with $q_c\sim 0.2$,
which will be removed in the results.

\begin{figure}[htb]
  \includegraphics[width=\linewidth]{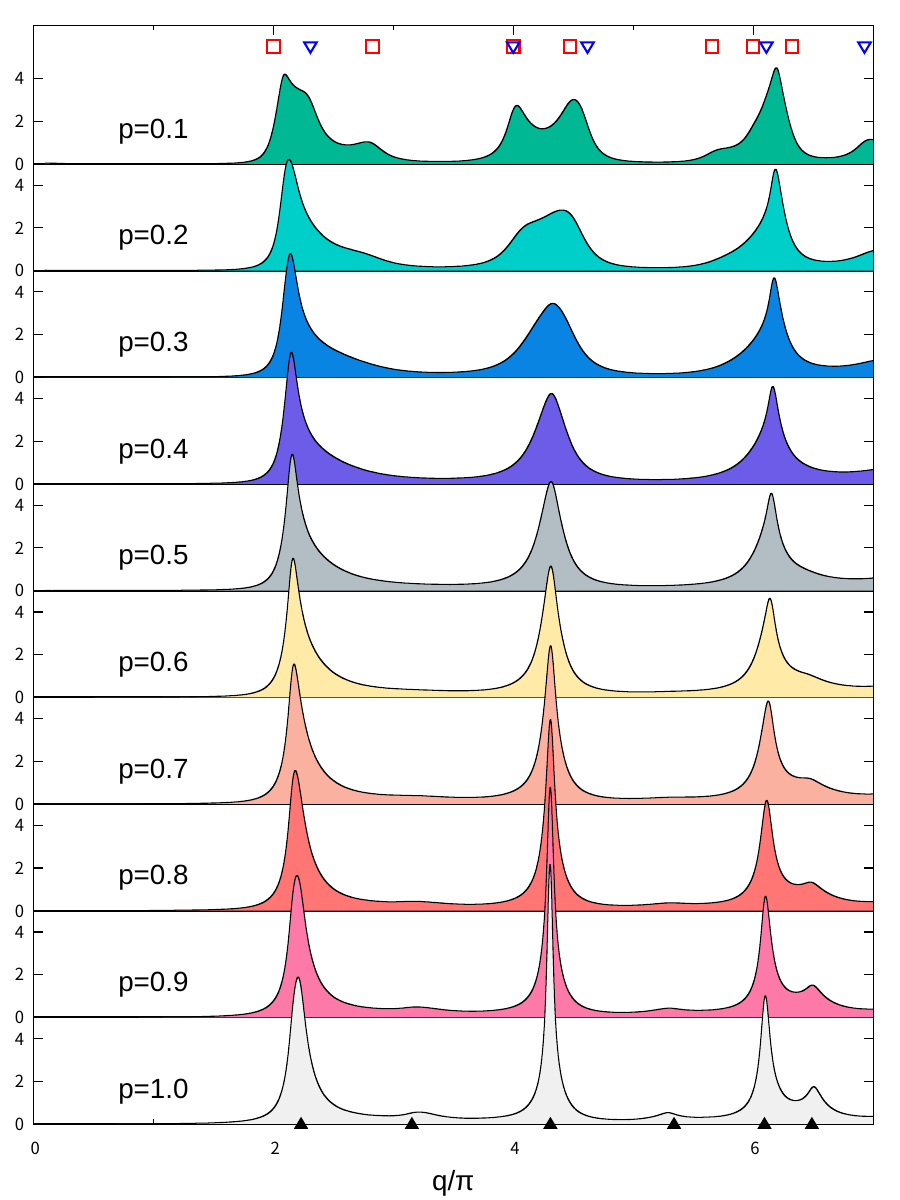}
  \caption{
    Structure factor $S(q)$ of the square-triangle tilings
    with $p=0.1, 0.2, \cdots$, and $1.0$ from the top to the bottom.    
    Open squares and triangles in the top panel indicate peak positions in the structure factors
    of the square and triangular lattices.
    Solid triangles in the bottom panel indicate those of the deterministic Stampfli quasiperiodic tiling.
  }
  \label{fig:Sq}
\end{figure}

Figure~\ref{fig:Sq} shows the structure factors for various values of $p$,
by examining more than million circular tilings.
We find several peaks in the structure factor for $p=0.1$.
These peak positions are close to those for the square and triangular lattices,
which are shown as open squares and triangles in Fig.~\ref{fig:Sq}.
This means that, for a small $p$,
the square-triangle tiling can be regarded 
as a mixture of square and triangular lattices,
which is consistent with the fact that
the large square and triangle domains appear in the tiling
[see Fig.~\ref{fig:tilings}(a)].
With increasing $p$, three peak structures mainly appear
in the range $0<q<7\pi$.
We find that these peak positions little depend on the parameter $p$
and are close to
those for the hyperuniform Stampfli quasiperiodic tiling,
which are shown as the solid triangles in Fig.~\ref{fig:Sq}.
In addition, we clearly find that two peaks
at $q/\pi \sim 4.3$ and $6.1$ become sharper
with increasing $p$.
This may imply that the tiling structure approaches
the deterministic quasiperiodic tiling,
whose structure factor is represented by a set of the delta-functions.
On the other hand, different behavior appears
in the first peak at $q/\pi \sim 2.2$
and its width little changes.
The detail of the first peak will be discussed later.

\begin{figure}[htb]
  \includegraphics[width=\linewidth]{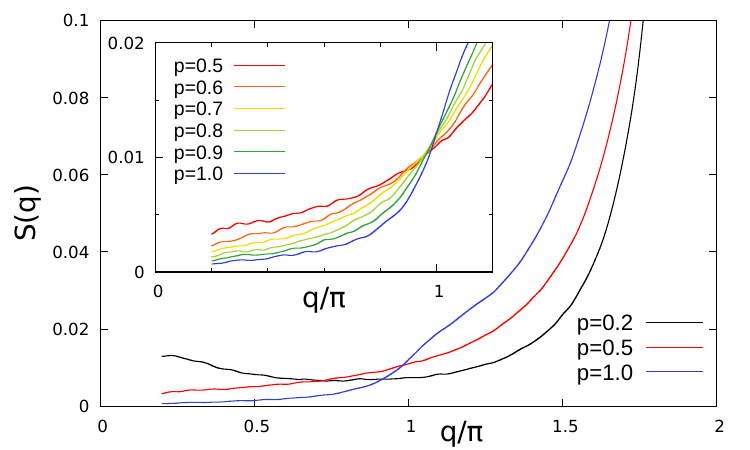}
  \caption{
    Structure factor $S(q)$ around $q=0$ obtained by combining 
    Eq.~(\ref{eq:sq}) with the Gaussian filter with $\sigma=0.07$.
  }
  \label{fig:Sq2}
\end{figure}
In the small $q$ region ($q/\pi\lesssim 2$),
the structure factor seems zero in this scale, as shown in Fig.~\ref{fig:Sq}.
However, this does not necessarily imply that
the randomly distributed square-triangle tilings obtained here are always hyperuniform.
To clarify hyperuniform properties,
we focus on the low $q$ region.
The magnified figure around $q=0$ is shown in Fig.~\ref{fig:Sq2}.
We find that in the case with $p\lesssim p_c$,
the structure factor tends to be finite for $q\rightarrow 0$.
This means that
the point configurations are not hyperuniform.
On the other hand, for $p\gtrsim p_c$,
the structure factor tends to decrease with decreasing $q$,
as shown in the inset of Fig.~\ref{fig:Sq2}.
This is consistent with the fact that the point configurations are hyperuniform.
We believe that $S(q)\sim q^\alpha$ with $q\rightarrow 0$
although the structure factor in the small $q(<q_c)$ region
could not be deduced quantitatively from our numerical simulations.

\begin{figure*}[htb]
  \includegraphics[width=\textwidth]{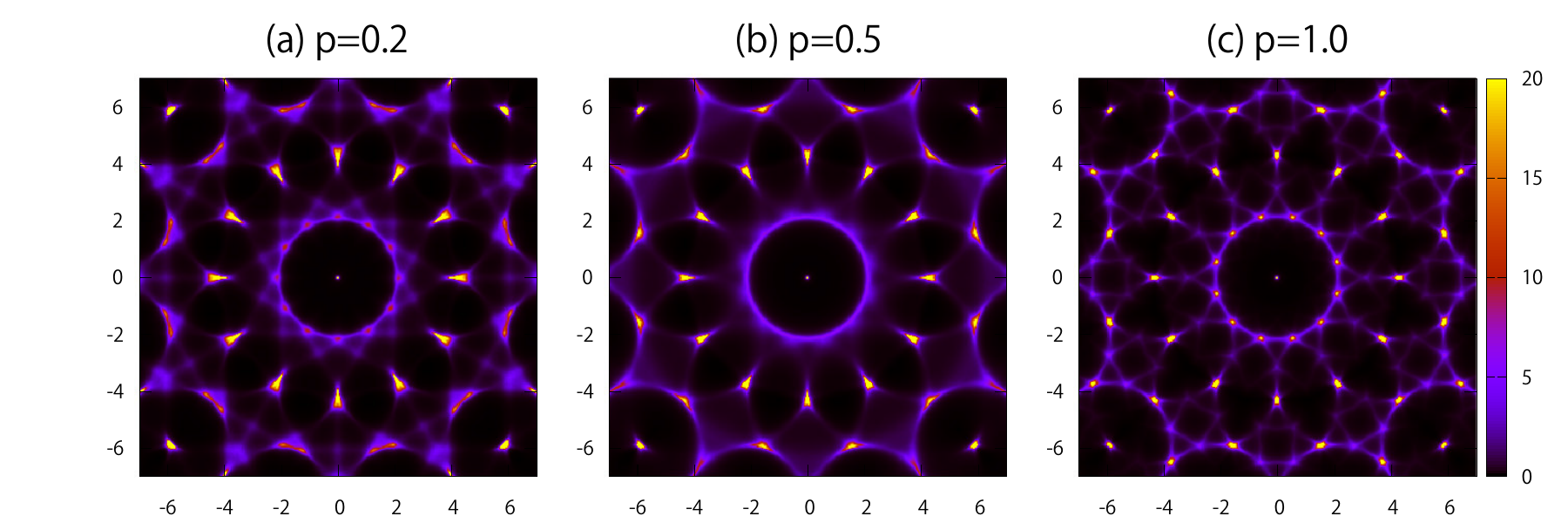}
  \caption{
    Structure factors of the square-triangle tilings with $R_{max}=300$
    when (a) $p=0.2$, (b) $p=0.5$, and (c) $p=1.0$. 
  }
  \label{Sqvec}
\end{figure*}
We also show the structure factor in the reciprocal space in Fig.~\ref{Sqvec}.
We find that many peaks appear in the structure factor and
these have dodecagonal symmetry.
Now, we focus on three kinds of main peaks
with $q/\pi\sim 2.2, 4.3$, and $6.1$.
One finds that the positions of the latter two peaks little depend
on $p$.
This implies that these common peaks are characteristic of
the dodecagonal tilings composed of the squares and triangles.
In fact, we find that similar peak structures appear
in the structure factor of the deterministic dodecagonal Stampfli quasiperiodic tiling.
On the other hand, different behavior appears
in the first peak around $q/\pi\sim 2.2$.
For $p=0.2$, 
the peak positions are characterized by the angle
$\theta=2n \pi/12$ with $n=0, 1, \cdots$, and $11$,
as shown in Fig.~\ref{Sqvec}(a).
Twelve peaks are also found in the case of $p=1.0$,
while their peak positions are characterized by the angle $\theta=(2n+1) \pi/12$,
as shown in Fig.~\ref{Sqvec}(c).
When $p$ is continuously varied,
the peak structures smear around $p=0.5$,
as shown in Fig.~\ref{Sqvec}(b).
We would like to emphasize that 
this crossover occurs at the hyperuniform-antihyperuniform transition point
$(p=p_c)$.
For $p=0.5$, $S({\bf q})$ has an isotropic structure in the long length
and dodecagonal symmetric structure in the small length.
It is interesting to clarify how the point configurations with two distinct properties
affect optical response, ground state properties, etc., which will be discussed in the future.

\section{Discussions}\label{discussion}
\begin{figure*}[htb]
  \includegraphics[width=0.65\linewidth]{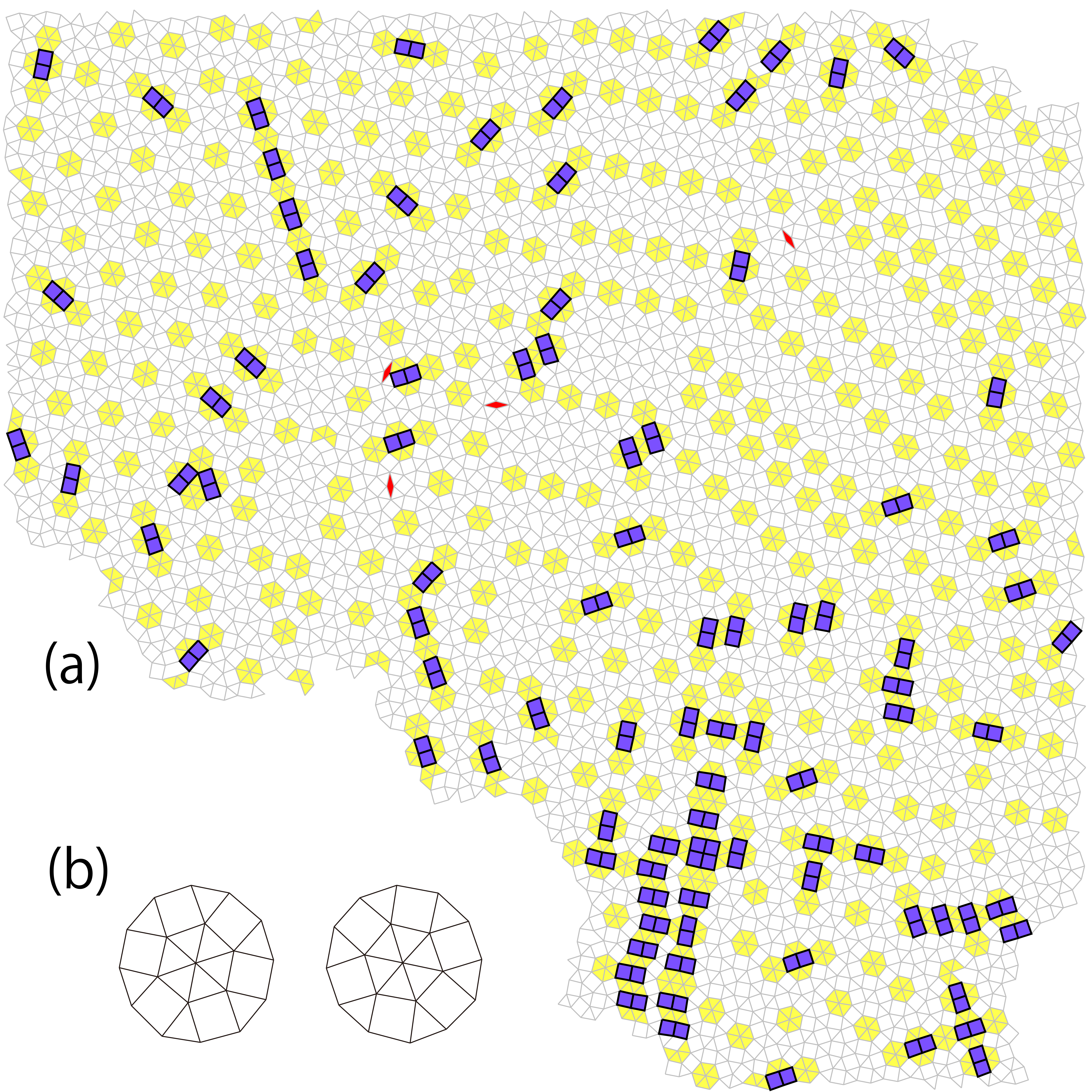}
  \caption{
    (a) Tiling structure observed in the Mn-Cr-Ni-Si alloy mainly 
    composed of squares and triangles~\cite{Ishimasa_2015}. 
    Bold purple squares represent multiple square assembly more than two, 
    while yellow triangles express domains composed of more than three consecutive triangles.
    Local structures around yellow hexagons correspond to those in (b).
    There are few red rhombuses. 
    (b) Local dodecagonal arrangements of triangles and squares.
  }
  \label{ex1}
\end{figure*}

We would like to comment on the relevance of the realistic materials
with dodecagonal symmetry.
Here, we examine their spatial structures, in particular,
the domain structure composed of the squares or triangles.
In general, in inorganic materials, there are stable local structures
depending on the atoms and their orbitals. 
Therefore, 
regular square and triangle structures are often realized and 
edge lengths of the square and triangle are almost fixed.
In the compounds, such regular squares and triangles nearly cover a two-dimensional sheet
and their structure factor is consistent with that for our model with a large $p$.
The structure of the Mn-Cr-Ni-Si alloy~\cite{Ishimasa_2015}, 
which has been observed by Cs-corrected high-resolution electron microscopy, 
is shown in Fig.~\ref{ex1}(a). 
In this case, the common edge length of the squares and triangles is 4.6$\AA$. 
They are well mixed, and tend to form a dodecagon presented in Fig.~\ref{ex1}(b). 
This dodecagon includes the hexagon at its center. 
There are two orientations for the hexagons. 
As a whole, neither large square nor triangle domains appear. 
We also find a few rhombuses as defects of the tilings. 
A similar structure has been observed in other compounds such as Ta$_{1.6}$Te~\cite{Conrad_1998}. 
The frequent appearance of the dodecagons is in contrast to our tiling with $p = 1.0$ [see Fig.~\ref{fig:tilings}(c)]. 
Therefore, the growth rule with additional conditions should be necessary to explain the realistic inorganic materials. 

\begin{figure*}[htb]
  \includegraphics[width=0.9\linewidth]{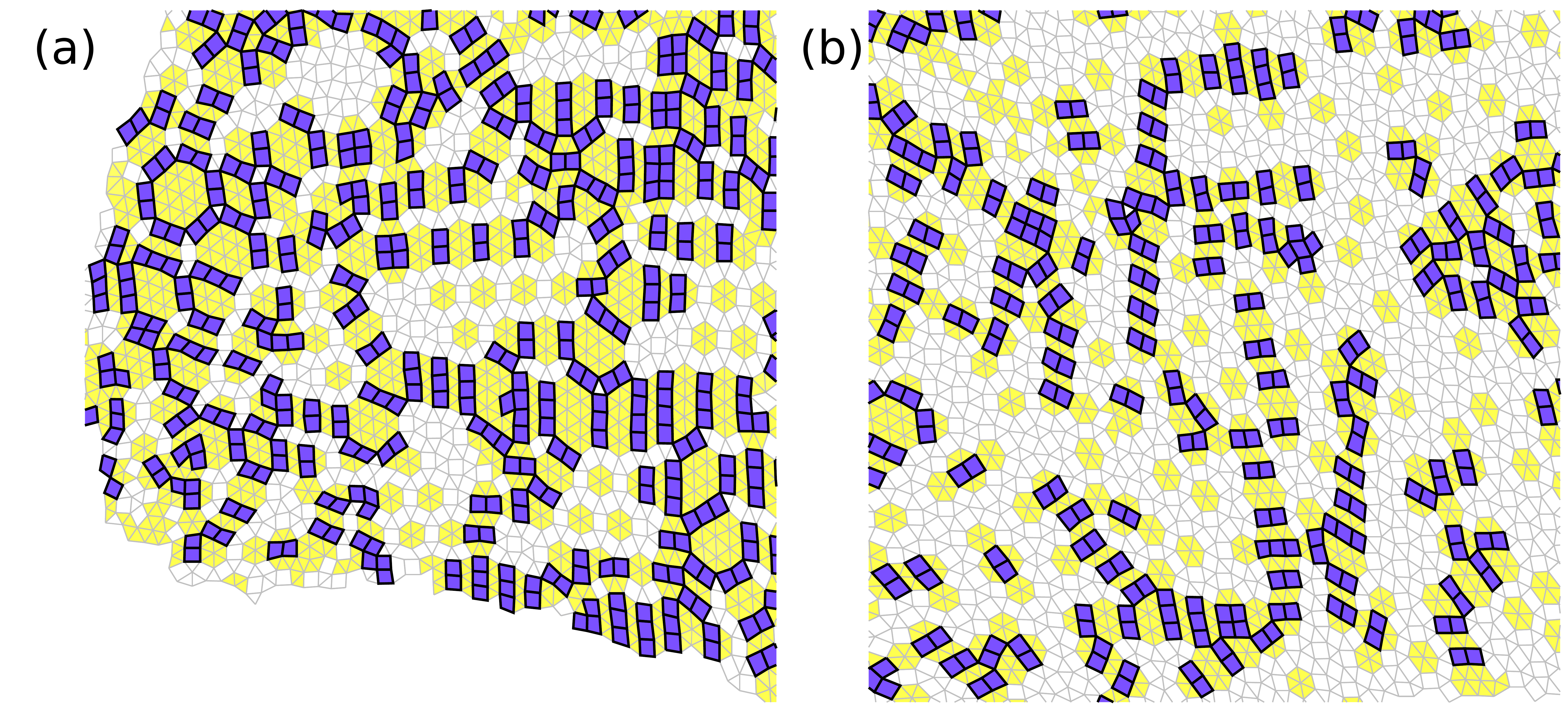}
  \caption{
    The tiling structures obtained by the multicomponent block polymer experiments. 
    Bold purple squares represent multiple square assembly more than two, 
    while yellow triangles express domains composed of more than three consecutive triangles.
    }
  \label{ex2}
\end{figure*}

In the multicomponent block polymer materials with
small composition distribution,
the system was found to form easily triangle-square random tiling
with dodecagonal symmetry~\cite{Matsushita}.
The resulting structure is similar to inorganic materials
though the side length of tiles is two order of magnitude larger.
Moreover, another type of triangle-square tiling composed of
relatively large triangle and square domains has been obtained
from the same complex polymer system, as shown in Fig.~\ref{ex2}(a).
Focusing on the density and space of the domains,
one notices that this experimentally observed tiling is similar to that
in Fig.~\ref{fig:tilings}(c) for a large $p$.
Another interesting image with smaller domain assembly is obtained as shown in Fig.~\ref{ex2}(b).
Here we find purple domains tend to be arranged in parallel, which is consistent with the simulated structure with $p=1$
[see Fig.~\ref{fig:tilings}(c)].

The results are also supported by the experimental results 
for the mesoporous silica~\cite{Xiao_2012,Wang_2020}. 
In Fig. 4(c) of Ref.~\cite{Wang_2020}, 
one can observe the crystal growth for the square-triangle tiling. 
Surprisingly, large square and triangle domains appear and 
some domains tend to be arranged in parallel.
This is consistent with our results with small $p$.
Systematic analysis for several CES/C$_{18-3-1}$ molar ratios has led
to a variety of square-triangle configurations although the system size is relatively small.
Although it is not clear how the CES/C$_{18-3-1}$ molar ratio is related to our control parameter $p$,
our growth rule proposed here should be an appropriate model
to grow the mesoporous silica as well as polymer crystal.

\section{Summary}\label{sec:summmary}
We have studied hyperuniform properties for the square-triangle tilings
generated by the growth rule.
The introduction of the probability $p$ in the growth rule
allows us to control the expansion of the square and triangle domains.
It has been clarified that when $p<p_c$,
the system should be regarded as the phase separation between square and triangular lattices 
and the variance of the point configurations
obeys the scaling law $\sigma^2\sim O(R^{2-\alpha})$ with $\alpha<0$.
Therefore, 
antihyperuniform point configurations are realized.
On the other hand, in the case with $p>p_c$,
the squares and triangles are spatially well mixed and
the point configurations belong to the hyperuniform class III
with the exponent $0<\alpha<1$.
We have also examined the structure factor using the two-point correlation functions.
It has been found that
the structure factor $S(q)$ with $q\sim 0$ should be finite for $p<p_c$,
while should be zero for $p>p_c$, which is consistent with the results
for the variance.
This difference in the long-range distances leads to
distinct dodecagonal peak structures for small wave numbers in the reciprocal space;
when $p<p_c \;(p>p_c)$,
the peaks in the small $q$ region
are characterized by the angle $2n \pi/12$ [$(2n+1)\pi/12$].
For $p=p_c$, the isotropic peak structure appears,
in contrast to the dodecagonal peak structure in large $q$ region, 
meaning that the system is quasicrystal-like in a short length scale 
while amorphous-like in a large length scale.

In this study, we have generated hyperuniform and antihyperuniform point configurations 
on the square-triangle tilings constructed by closely packing the plane with squares and triangles.
It is also intriguing to consider toy models defined by the connectivity of the vertices.
Results for the tight-binding model and lattice vibration are shown in Appendix~\ref{toy}.
Furthermore, it is known that the hyperuniformity is useful to discuss a spatial distribution 
of scalar quantities on a quasiperiodic tiling~\cite{Fuchs,Sakai_2019,Sakai_2022}.
Therefore, it is interesting to consider strong electron correlations in the square-triangle tilings
to discuss the spatial distribution of the order parameters,
which should be related to the superconductivity in Ta$_{1.6}$Te.
Another application is the photonic crystal.
The photonic crystal with the disordered hyperuniform point configurations,
which are not based on the square and triangle tiles,
has been created, where the isotropic band gap is realized~\cite{Florescu_2009,Florescu_2009_2,Man_2013,Gkantzounis_2017}.
On the other hand, a large optical band gap has also been realized in
the photonic crystal on the dodecagonal Stampfli quasiperiodic tiling~\cite{Zoorob_2000}.
Therefore, a larger and more isotropic photonic gap is expected
in the photonic crystal on our tilings,
which will be discussed in the future.

\begin{acknowledgments}
  We would like to thank K. Edagawa, N. Fujita, M. Imp\'eror-Clerc, and Y. Ishii for valuable discussions.
  Parts of the numerical calculations were performed
  in the supercomputing systems in ISSP, the University of Tokyo.
  This work was supported by Grant-in-Aid for Scientific Research from
  JSPS, KAKENHI Grant Nos. JP22K03525 (A.K.).
\end{acknowledgments}

\begin{appendix}

\section{Growth of the tilings}\label{G}

We use a local growth rule,
which has been proposed in Sec.~\ref{rule}, to generate
the square-triangle tiling.
In general, when the tiling is generated by means of the growth rule,
the overlaps of the tiles are sometimes encountered.
One of the main reasons is that the tiling grows with a significantly curved interface.
To avoid the mismatch, we consider the vertex closest to the origin
and attach the pair of tiles
when its arrangement is randomly determined.
Then, we generate the large square-triangle tiling with $R\sim O(10^3)$,
where the mismatch rarely occurs.
Up to now, these rare events could not be removed completely.
Therefore, we discuss hyperuniform properties
after the tilings without mismatches are obtained.

We note that the local tiling structure
depends on the initial structure.
when one makes use of the above growth rule.
To obtain the tiling independent of an initial structure,
we use the following method.
First, we consider a certain circular initial tiling with a radius $R_0^{(0)}$ and grow
the tiling with a radius $R_1^{(0)} (>R_0^{(0)})$.
We then choose a certain circular domain with a radius $R_0^{(1)}$ in the tiling
and regard it as an initial tiling.
After this procedure is repeated twice,
we obtain the initial circular tiling.
Then, we generate the large circular tiling with $R\sim 1000$ and
randomly choose the tiling with $R\sim 100$ inside.
We believe that the obtained tilings less depend on 
the distance and direction from the origin, and thereby
the main results do not change qualitatively.


\section{Perpendicular space}\label{Ap:perp}

The square-triangle tiling obtained by the growth rule
is distinct from the deterministic Stampfli quasiperiodic tiling~\cite{Stampfli,Hermisson_1997}.
This can be clearly explained by considering the vertex distribution
in the perpendicular space.
This space is perpendicular to the physical one
in the high-dimensional description of a quasicrystalline structure and
the positions in the perpendicular space have one-to-one correspondence
with the positions in the physical one.
Each vertex in the tiling is described by a four-dimensional lattice
point $\vec{n} = (n_0, n_1,n_2,n_3)$, labeled with integers $n_m$,
where the lattice is spanned by four fundamental translation vectors.
The coordinates of the tiling are the projections onto
the two-dimensional physical space, as
\begin{align}
  {\bf r} &= \sum_{i=0}^3 n_m {\bf e}_m,
\end{align}
where ${\bf e}_m[=(\cos m\pi /6 , \sin m\pi/6)]$ is the projected vector.
We can then project the points onto the two-dimensional perpendicular space
as
\begin{align}
  {\bf r}^\perp & = \sum_{i=0}^3 n_m {\bf e}^\perp_m,
\end{align}
where ${\bf e}_m^\perp={\bf e}_m$ for $m=0,2$ and
${\bf e}_m^\perp=-{\bf e}_m$ for $m=1,3$.

\begin{figure}[htb]
  \includegraphics[width=\linewidth]{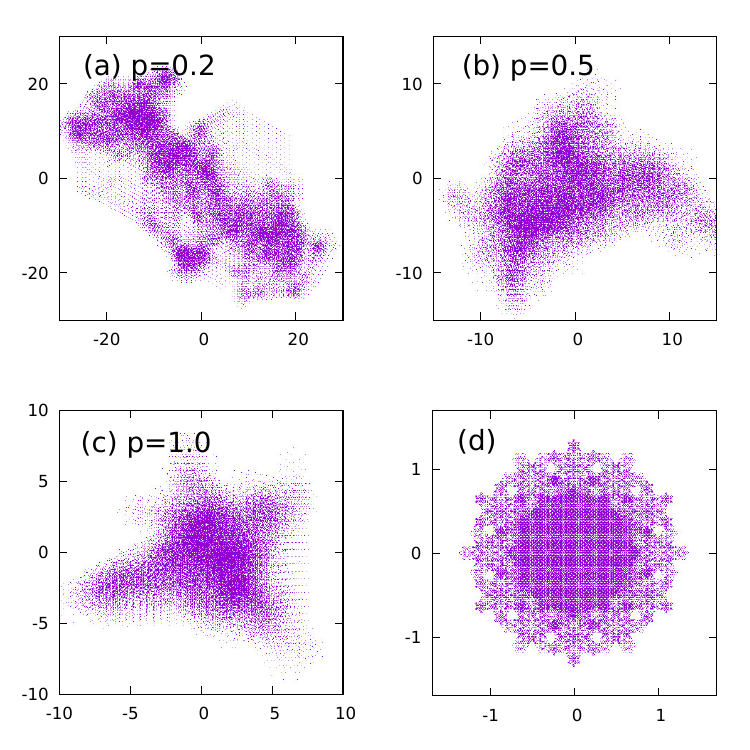}
  \caption{
    Vertex distributions of the square-triangle tiling with $R_{max}=100$
    with (a) $p=0.2$, (b) $p=0.5$, and (c) $p=1.0$
    in the perpendicular space.
    (d) represents the vertex distribution of the dodecagonal Stampfli quasiperiodic tiling
    with $R_{max}=100$.
    Notice the difference in the scales.
  }
  \label{perp}
\end{figure}

It is known that for the deterministic Stampfli quasiperiodic tiling,
the domain structure does not appear in the perpendicular space,
but fractal behavior appears with twelve-fold rotational symmetry~\cite{Baake_1992}.
On the other hand, it is known that
no rotational symmetry appears in the random tilings~\cite{HenleyBook}.
Here, we show in Fig.~\ref{perp}
the vertex distributions in the perpendicular space
for the square-triangle tiling for one sample with $R_{max}=100$.
For the deterministic Stampfli quasiperiodic tiling,
fractal behavior with rotational symmetry
appears in the perpendicular space.
This means that the squares and triangles are mixed in the physical space
and thereby the vertex distributions are limited within a certain region in the perpendicular space.
On the other hand, the square-triangle tiling obtained in our paper
includes the square and triangle domains.
In this case, a certain vertex is represented by
the four-dimensional vector with large $|n_m|$,
in contrast to the Stampfli quasiperiodic tiling.
This leads to a relatively large distribution without rotational symmetry
in the perpendicular space.
Furthermore, we find the larger distribution
in the smaller $p$ case.
This is consistent with the fact that, increasing $p$,
the point configuration approaches the deterministic quasiperiodic one
belonging to the hyperuniform class I.

\begin{figure}[htb]
  \includegraphics[width=0.8\linewidth]{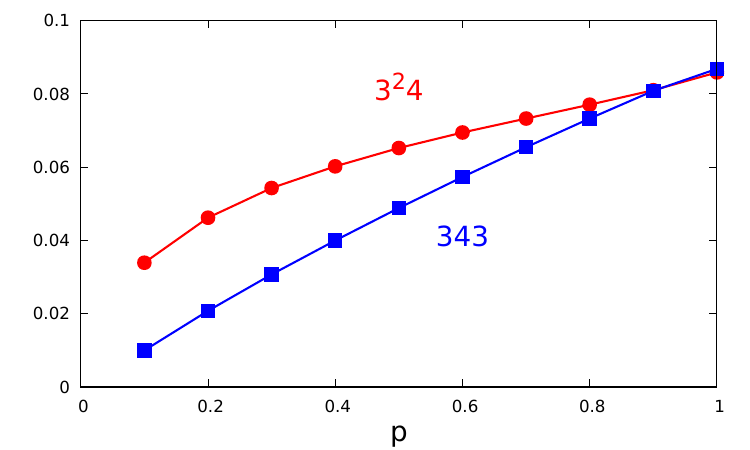}
  \caption{
  Ratios for $3^24$ and $343$ vertices which appear on the surface 
  when the tiling is growing.
  The data is obtained, by examining a thousand independent circular tilings with $R=300$.
}
  \label{prob}
\end{figure}

We would like to discuss the relationship with the random tilngs in Ref.~\cite{HenleyBook}.
Since our tilings are generated by means of the growth rule,
it is difficult to make a direct comparison with those tilings.
In the perpendicular space, the vertex structures in our tilings
lack rotational symmetry, as discussed above.
The number of the allowed configurations should be related to
the number of times during the growth process where the local arrangements of tiles are randomly chosen.
When the tilings are growing, 
one sometimes meets $3^24$ and $343$ vertices on the surface, 
randomly chooses the arangement of square and triangle tiles, and attaches the pair of tiles,
as discussed in Sec.~\ref{rule}.
Figure~\ref{prob} shows the ratios of the $3^24$ or $343$ vertices 
met during the growth process to the total number of sites.
We find that each ratio is always finite for $p\neq 0$.
This implies that the number of allowed configurations grows exponentially.
These facts are consistent with the characteristics of random tilings described in Ref.~\cite{HenleyBook}.

\section{Structure factor}\label{A}
  
\begin{figure}[htb]
  \includegraphics[width=\linewidth]{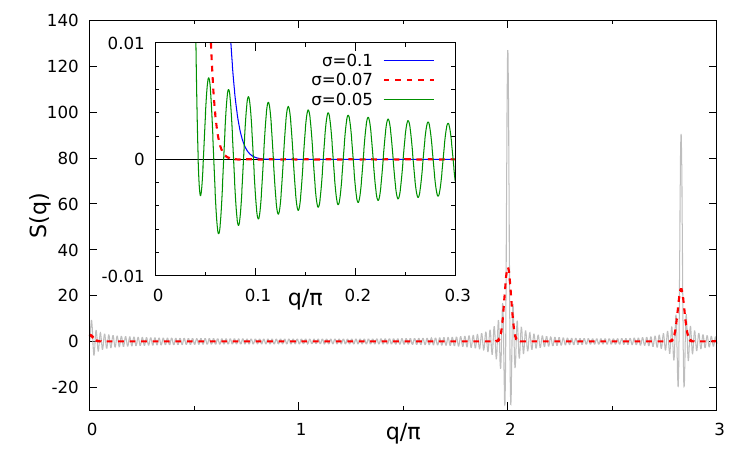}
  \caption{
    Solid (dashed) line represents the bare (improved) structure factor of the square lattice.
    Inset shows the width dependence of the improved structure factor around $q\sim 0$. 
  }
  \label{square}
  \end{figure}

We focus on the structure factor in the vicinity of $q=0$,
which is important for discussing hyperuniform properties.
In our calculations, the structure factor is evaluated
by the histogram
\begin{align}
  h(r)=\frac{1}{N}\sum_{i\neq j}\delta_{r,r_{ij}}=\sum_k\omega_k\delta_{r,r_k},
\end{align}
where $\omega_k$ is the weight of the two-point distance $r_k$.
Since the range of the histogram is restricted
to the region with $r_k<R_{max}$,
oscillation behaviour appears in the structure factor,
which makes it difficult to discuss low $q$ behavior
in the structure factor.
To overcome this, we use the Gaussian filter.
The improved structure factor is given in terms of the Gauss function as
\begin{eqnarray}
  \bar{S}(q)&=&\int dq' S(q')f(q-q'),\\
  f(q)&=&\frac{1}{\sqrt{\pi}\sigma}\exp\left[-\left(\frac{q}{\sigma}\right)^2\right],
\end{eqnarray}
where $\sigma$ is the width of the Gaussian.
The bare structure factor and the modified one
for the square lattice are shown in Fig.~\ref{square}.
Both are obtained by the histogram with $R_{max}=100$.
One finds that large oscillation behavior appears in the bare structure factor
when $q\sim 0$ and $\sim 2\pi$.
Therefore, hyperuniform properties should be difficult to discuss
in terms of the bare structure factors $S(q)$.
On the other hand, the oscillation is suppressed when
the Gaussian filter with $\sigma\gtrsim 0.07$ is used,
as shown in Fig.~\ref{square}.
The obtained structure factor is almost zero in a small $q$ region,
which is consistent with
the fact that it has a first delta function-like peak at $q=2\pi$
and is zero with $0<q<2\pi$.
Then, we can say that the correct results are reproduced
except for the tiny region $q< q_c$, with $q_c\sim 0.1$.
Since the cutoff $q_c$ depends on $R_{max}$, 
this finite size effect can not be completely removed
in the tiny region with $q<q_c$.
Nevertheless, one can deduce the structure factor $S(q)$ with $q\rightarrow 0$,
extrapolating the curve.

\section{Toy models on the square-triangular tiling}\label{toy}

\begin{figure}[htb]
  \includegraphics[width=\linewidth]{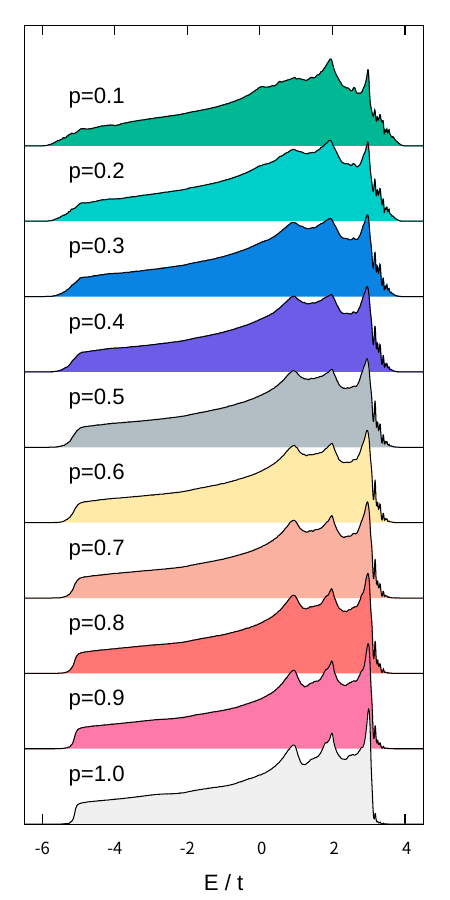}
  \caption{
    Density of states for the tight-binding model
    on the square-triangle tiling.
  }
  \label{dos}
  \end{figure}

Here, we consider the tight-binding model on the square-triangle tiling.
The model Hamiltonian is given as,
\begin{eqnarray}
  H=-t\sum_{(i,j)} c_i^\dag c_j,
\end{eqnarray}  
where $(i,j)$ represents the nearest neighbor pair of the vertices and
$c_i(c_i^\dag)$ annihilates (creates) a particle at the $i$th site.
$t$ is the hopping integral.
  Considering the independent 30 circular tilings with $\sim 40,000$ vertices,
  diagonalizing the corresponding Hamiltonians,
  we obtain the density of states,
  as shown in Fig.~\ref{dos}.
We find that the band edge becomes sharper with increasing $p$
although no drastic change appears in the density of states
of this simple model.
It is an interesting problem to clarify how the Anderson localization
appears in the square-triangle tiling,
by evaluating how localized each eigenstate is.

It is also instructive to consider the lattice vibration~\cite{Gkantzounis_2017}.
When the atoms are located at the vertices on the square-triangle tiling,
the classical motion of equation is given by Hooke's law as
\begin{eqnarray}
  m\frac{d^2{\bf r}_i}{d t^2}&=&m\omega_0^2\sum_\delta\left({\bf r}_i-{\bf r}_{i+\delta}\right)\\
  &=&m\omega_0^2\left(z_i{\bf r}_i-\sum_\delta{\bf r}_{i+\delta}\right)
\end{eqnarray}
where $m$ is the mass of the atom, $\omega_0$ is the frequency characteristic of 
the vibration between nearest neighbor sites, and
$z_i$ is the coordination number for the $i$th site.
If the coordination number takes the same number $z$, the eigenequation is essentially
the same as that for the tight-binding model and
the eigenvalue is given by $\omega_i/\omega_0 = \sqrt{E_i/t+z}$,
where $E_i$ and $\omega_i$ are the $i$th eigenvalues of the tight-binding and vibration models,
respectively.
In the square-triangle tiling, the coordination number takes $z_i=4, 5, 6$,
depending on the vertex (see Fig.~\ref{fig0}).
Therefore, the density of states for the vibration model is not obtained
using the eigenvalues of the corresponding tight-binding model.
\begin{figure}[htb]
  \includegraphics[width=\linewidth]{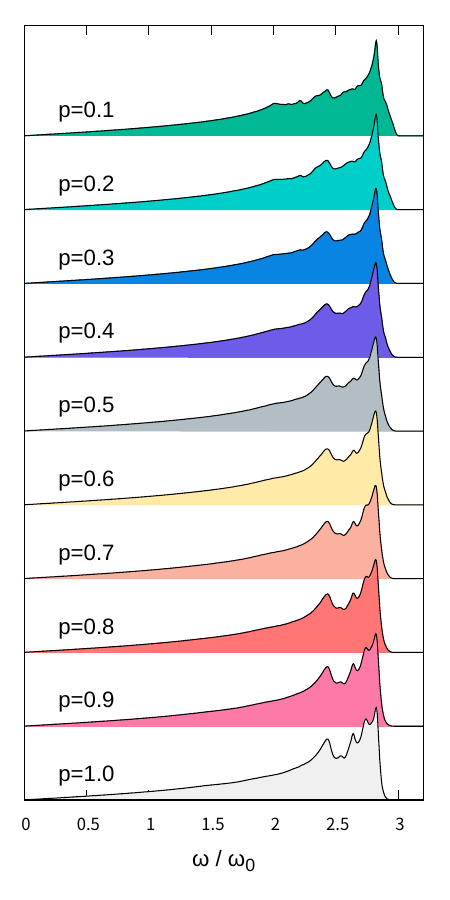}
  \caption{
    Density of states for the phonon dispersion
    on the square-triangle tiling.
  }
  \label{phonon}
  \end{figure}
Figure~\ref{phonon} shows the density of states for the phonon dispersion,
which is obtained by averaging the results for the independent 30 circular tilings with $\sim 40,000$ vertices.
We find that the overall structure has little changes.
In the low energy region, the density of states is proportional to $\omega$,
indicating the presence of the acoustic phonons.
The velocity of the acoustic phonon,
which is given by the slope of the density of states
in the low $\omega$ region, remains nearly unchanged.
However, the band edge around $\omega/\omega_0\sim 3$
becomes sharper with increasing $p$,
similar to the behavior in the tight-binding model.
In the tilings, the squares and triangles are always mixed, 
and thereby it is hard to examine tight-binding and vibration models 
when triangles are introduced into a square lattice~\cite{FLO}. 
This intersting problem will be discussed elsewhere. 

\begin{figure}[htb]
  \includegraphics[width=\linewidth]{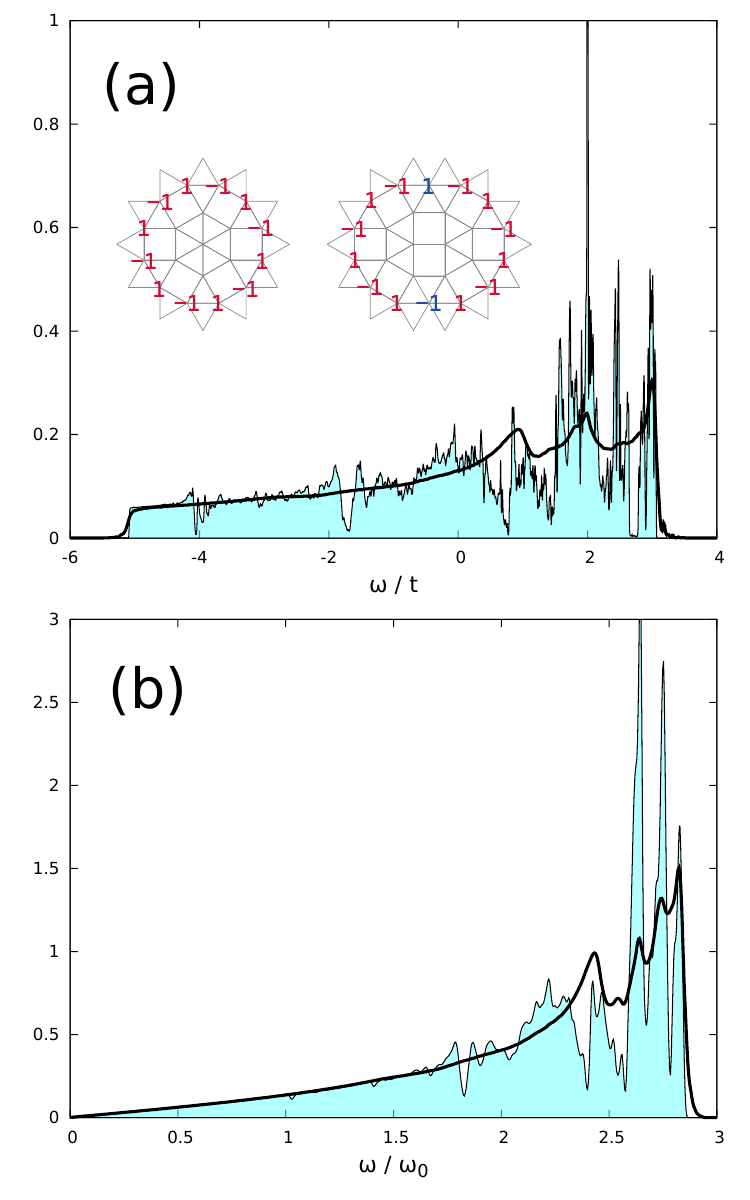}
  \caption{
    Densities of states for (a) the tight-binding and (b) vibration models.  
    Thin and bold lines represent the results 
    for the Stampfli quasiperiodic tiling and square-triangle tiling with $p=1$, respectively.
    Inset of (a) shows two kinds of the confined states with $\omega=2t$. 
    The red (blue) numbers at the vertices with $z_i=5 (6)$ 
    represent the amplitudes of confined states.
    }
  \label{combine}
  \end{figure}

Finally, we compare these results with those for the deterministic Stampfli quasiperiodic tiling.
The densities of states for the tight-binding and vibration models are
obtained by diagonalizing the Hamiltonians with $305,797$ vertices.
The results are shown in Fig.~\ref{combine}.
In contrast to the results for the square-triangle tilings,
sharp peak structures appear in the density of states.
We find that, in the tight-binding model, there exists the delta function like peak at $\omega=2t$.
This originates from the existence of the confined states~\cite{Kohmoto_Sutherland_1986,Fujiwara_1988,Arai_1988}.
Two of the confined states with $\omega=2t$ are explicitly shown in the inset of Fig.~\ref{combine}(a).
Since the deterministic Stampfli quasiperiodic tiling is constructed by the substitution rule~\cite{Hermisson_1997},
any finite part of the lattice repeats itself within a certain distance.
Therefore, the eigenstates with $\omega=2t$ are macroscopically degenerate,
leading to the delta function peak in the density of states.
There exist such lattice structures even in the square-triangle tiling 
similar to the quasiperiodic tiling.
However, these are sparsely distributed and 
thereby one can see the broader peak rather than the delta function peak in the density of states. 

For the vibration model, 
the delta function like peak appears at $\omega = \sqrt{7}\omega_0$,
as shown in Fig.~\ref{combine}(b).
When the vibration mode involves only the vertices with the same coordination number,
the eigenequation is essentially the same as that for the tight-binding model, as discussed above.
Hence, the state shown left in the inset of Fig.~\ref{combine}(a) is the eigenstate 
of the vibration model, while the right one is no longer the eigenstate. 
As a result, the peak weight for the confined state is higher in the tight-binding model
than in the vibration model.
Although localized phonon modes macroscopically exist in the deterministic tiling,
the corresponding peak structure smears in the square-triangle tiling with $p=1$,
as similar to the tight-binding model.

\end{appendix}

\nocite{apsrev42Control}
\bibliographystyle{apsrev4-2}
\bibliography{./refs}

\end{document}